\documentclass[iop]{emulateapj}

\usepackage{graphicx} 
\usepackage{amsmath}
\usepackage{appendix}
\usepackage{amssymb}
\usepackage{aasmacros}
\usepackage{natbib}
\usepackage{latexsym}
\usepackage{longtable}
\usepackage{threeparttable} 
\usepackage{subfigure} 
\usepackage[figoff]{figcaps}
\usepackage{ccaption}
\usepackage{hyperref}
\captiondelim{ }
\usepackage[usenames,dvipsnames]{color}

\begin{document}

\pagestyle{plain}
\title{Near-Earth Binaries and Triples: Origin and Evolution of Spin-Orbital Properties}
\author{Julia Fang\altaffilmark{1} and Jean-Luc Margot\altaffilmark{1,2}}

\altaffiltext{1}{Department of Physics and Astronomy, University of California, Los Angeles, CA 90095, USA}
\altaffiltext{2}{Department of Earth and Space Sciences, University of California, Los Angeles, CA 90095, USA}

\begin{abstract}

In the near-Earth asteroid population, binary and triple systems have
been discovered with mutual orbits that have significant
eccentricities as well as large semi-major axes.  All known systems
with eccentric orbits and all widely-separated primary-satellite pairs
have rapidly rotating satellites. Here we study processes that can 
elucidate the origin of these spin-orbital properties.
Binary formation models based on rotational fissioning can reproduce
asynchronous satellites on orbits with high eccentricities and a wide
range of separations, but do not match observed properties.
We explore whether any evolutionary mechanisms can
link the spin and orbital parameters expected from post-fission
dynamics to those observed today.  We investigate four processes:
tidal torques, radiative perturbations (BYORP), close planetary
encounters, and Kozai oscillations.
We find that a combination of post-fission dynamics and tidal
evolution can explain nearly all the spin-orbit properties in a 
sample of nine well-characterized near-Earth binaries and triples.  
The other mechanisms may act but are not required to explain the
observed data. Lastly, we describe evolutionary pathways between
observed spin-orbital states including synchronous and circular,
asynchronous and circular, and asynchronous and eccentric
configurations.

\end{abstract}
\maketitle

\section{Introduction} \label{intro}

The study of near-Earth asteroids (NEAs) and main belt asteroids
(MBAs) with satellites can yield important information about their
fundamental physical properties as well as their formation and
evolution \citep{merl02,prav06}. In these multi-component systems,
analysis of the relative positions between components can 
quantify their mutual orbits as well as the total mass of the system.  In
the near-Earth population, binary and triple asteroids are typically
discovered by radar (24 out of 37 as of October 2011) during a close
approach to Earth, which can provide detailed physical and orbital
information about the system. Studies by \citet{marg02} and
\citet{prav06} have determined that approximately 15\% of NEAs larger
than 200 meters in diameter have satellites. Due to the $d^{-4}$
dependence in the return signal (where $d$ is the distance to the
target), radar observations have not identified MBA systems so far.
Small MBA systems have generally been discovered through light curve
observations whereas larger MBA systems have typically been discovered
with adaptive optics observations.

Orbital solutions of NEA systems indicate that some of their
satellites possess unexplained spin-orbital properties including
asynchronous\footnote{In this paper, binaries with an absence of
spin-orbit synchronism are called {\em asynchronous binaries}.
Binaries with a secondary spin period synchronized to the mutual orbit
period are called {\em synchronous binaries}.  Binaries with both
primary and secondary spin periods synchronized to the mutual orbit
period are called {\em doubly synchronous binaries}.  Most NEA
binaries are {\em synchronous}. Note that our terminology is different
from that of \citet{prav07}, who use the term ``asynchronous binaries"
for binaries with spin-orbit synchronization. If generalization to
systems with more than one satellite is needed, we affix the terms
{\em synchronous} and {\em asynchronous} to the satellites being
considered.} rotation, eccentric orbits, and wide separations from their
primaries. All known satellites with semi-major axes larger than 7
primary radii are asynchronously rotating and all known satellites 
with eccentricities greater than 0.05 are asynchronously rotating,
suggesting that these properties have a common origin.
Accordingly, we seek an explanation for these observed spin-orbital
characteristics by examining if any evolutionary processes can lead
to the observed data.

\def\arraystretch{1.4}
\begin{deluxetable*}{lrrrrrrrrr}
\tablecolumns{10}
\tablecaption{Well-Characterized Near-Earth Binaries and Triples \label{nea}}
\startdata
\hline \hline
\multicolumn{1}{c}{System} &
\multicolumn{1}{c}{$R_p$} & 
\multicolumn{1}{c}{$M_p$} & 
\multicolumn{1}{c}{$R_s$} & 
\multicolumn{1}{c}{$M_s$} & 
\multicolumn{1}{c}{$a$} &
\multicolumn{1}{c}{$e$} & 
\multicolumn{1}{c}{$a/R_p$} &
\multicolumn{1}{c}{CA Distance} & 
\multicolumn{1}{c}{MOID} \\
\multicolumn{1}{c}{} &
\multicolumn{1}{c}{(km)} & 
\multicolumn{1}{c}{(kg)} & 
\multicolumn{1}{c}{(km)} & 
\multicolumn{1}{c}{(kg)} & 
\multicolumn{1}{c}{(km)} &
\multicolumn{1}{c}{} & 
\multicolumn{1}{c}{} &
\multicolumn{1}{c}{(AU)} & 
\multicolumn{1}{c}{(AU)} \\
\hline
(185851) 2000~DP107\footnotemark[1]	& 0.40 	& 4.38 $\times$ $10^{11}$	&	0.150 & 2.19 $\times$ $10^{10}$ 		& 2.62 $\pm$ 0.162 & 0.01$_{-0.01}^{+0.015}$   & 6.6  	& 0.058 in 2008 & 0.015  \\
(66391) 1999~KW4\footnotemark[2] 	& 0.66	& 2.35 $\times$ $10^{12}$ 	&	0.226 & 1.35 $\times$ $10^{11}$ 	  & 2.55$_{-0.01}^{+0.03}$ & 0.008$_{-0.008}^{+0.012}$ & 3.9 & 0.089 in 2002 & 0.013  \\
(276049) 2002~CE26\footnotemark[3] 	& 1.75 	& 2.17 $\times$ $10^{13}$ 	&	0.150 & 1.37 $\times$ $10^{10}$ 		& 4.87$_{-0.12}^{+0.28}$ & 0.025$_{-0.006}^{+0.008}$   & 2.8  	& 0.102 in 2004 & 0.100 \\
*2004~DC\footnotemark[4] 		& 0.17 	& 3.57 $\times$ $10^{10}$ 	&	0.030 	& 1.96 $\times$ $10^{8\ }$ 		&  0.75$_{-0.05}^{+0.04}$ & 0.30$_{-0.04}^{+0.07}$    & 4.4  	& 0.026 in 2006 & 0.009 \\
*(164121) 2003~YT1\footnotemark[5] 	& 0.55 	& 1.89 $\times$ $10^{12}$ 	&	0.105 	& 1.32 $\times$ $10^{10}$ 		& 3.93$_{-0.13}^{+1.47}$ & 0.18$_{-0.01}^{+0.02}$ 	  	& 7.1  	& 0.073 in 2004 & 0.002 \\
(65803) Didymos\footnotemark[6] 	& 0.40 	& 5.24 $\times$ $10^{11}$ 	&	0.075 	& 3.45 $\times$ $10^{9\ }$ 		& 1.18$_{-0.02}^{+0.04}$ & 0.04$_{-0.04}^{+0.05}$  	  & 3.0  	& 0.048 in 2003 & 0.04 \\
*(35107) 1991~VH\footnotemark[7] 	& 0.60 	& 1.40 $\times$ $10^{12}$ 	&	0.240 & 8.93 $\times$ $10^{10}$ 		& 3.26$_{-0.04}^{+0.03}$ & 0.06 $\pm$ 0.02    & 5.4  	& 0.046 in 2008 & 0.026 \\
(153591) 2001~SN263 \#1\footnotemark[8]& 1.30 	& 9.17 $\times$ $10^{12}$ 	&	0.230 & 9.77 $\times$ $10^{10}$ 	  & 3.80$_{-0.02}^{+0.01}$   & 0.016$_{-0}^{+0.005}$ 	& 2.9  & 0.066 in 2008 & 0.049 \\
*(153591) 2001~SN263 \#2\footnotemark[8]& 1.30 & 9.17 $\times$ $10^{12}$ 	&	0.530 & 2.40 $\times$ $10^{11}$  	  & 16.63$_{-0.38}^{+0.39}$ & 0.015$_{-0.010}^{+0.022}$ & 13   & 0.066 in 2008 & 0.049 \\
(136617) 1994~CC \#1\footnotemark[9]& 0.31  & 2.59 $\times$ $10^{11}$ 	&	0.057 	& 5.80 $\times$ $10^{9\ }$ 	& 1.73 $\pm$ 0.02 & 0.002$_{-0.002}^{+0.009}$	& 5.6  & 0.017 in 2009 & 0.016 \\
*(136617) 1994~CC \#2\footnotemark[9]& 0.31  & 2.59 $\times$ $10^{11}$ 	&	0.040 	& 9.10 $\times$ $10^{8\ }$ 		& 6.13$_{-0.12}^{+0.07}$ & 0.19$_{-0.022}^{+0.015}$ & 20  & 0.017 in 2009 & 0.016 
\enddata

\tablenotetext{}{Well-characterized NEA binaries and triples as
  defined in this paper have some known physical and orbital
  properties: approximate component sizes (primary radius $R_p$,
  secondary radius $R_s$), masses (primary mass $M_p$, secondary mass
  $M_s$), semi-major axis $a$, and eccentricity $e$.  Known
  asynchronous satellites are marked with a *.  This table shows
  nominal values adopted for this study as well as plausible
  1$-\sigma$ uncertainties for $a$ and $e$. Uncertainties for sizes
  are roughly $\sim$20\% and for masses are roughly
  $\sim$10\%. Methods for obtaining parameters and uncertainties are
  described in the text (Section \ref{sample}).  Also shown here is
  close approach (CA; $\lesssim$ 0.1 AU) data, including the most
  recent approach at the time of radar observation and the current
  MOID (see text) with Earth, given by the JPL Small-Body
  Database. For the triple systems, 2001 SN263 and 1994 CC, we list
  the inner satellite first and the outer satellite second.}
\footnotetext[1]{\citet{marg02}} \footnotetext[2]{\citet{ostr06}}
\footnotetext[3]{\citet{shep06}} \footnotetext[4]{\citet{tayl08b}}
\footnotetext[5]{\citet{nola04}}
\footnotetext[6]{\citet{benn10}}
\footnotetext[7]{\citet{marg08,prav06}}
\footnotetext[8]{\citet{nola08,fang11}}
\footnotetext[9]{\citet{broz11}, \citet{fang11}}
\end{deluxetable*}

Previous attempts to investigate the observed properties of NEA
systems include tidal evolution as a mechanism for eccentricity
excitation or de-excitation \citep[][and references therein]{tayl11}.
Other studies have described the orbital evolution of small
binary asteroids by the binary YORP (BYORP) effect
\citep{cuk05,cuk07}. BYORP is caused by the asymmetric re-radiation of
light by an irregularly shaped secondary in synchronous rotation with
its primary, and this effect can cause orbital migration and an
increase or decrease of the mutual orbit's eccentricity
\citep{gold09,cuk10,mcma10b,mcma10a,stei11}.  An alternative process
to modify an orbit's semi-major axis and eccentricity is through close
scattering encounters by terrestrial planets, as described in a
companion paper by \citet{fang11i} for binaries and by \citet{fang11}
for triple systems. Another possibility is that NEA binaries can have
their eccentricities excited through Kozai oscillations
\citep{koza62}.

In this paper, we examine all of these proposed evolutionary processes
to find a coherent model that can explain the observed spin-orbital
characteristics of satellites in NEA systems. In the remainder of this
section, the current population of well-characterized NEA binaries and
triples is presented in Section \ref{sample}, relevant lifetimes are
defined in Section \ref{lifetimes}, and binary and triple formation is
introduced in Section \ref{formation}. Then, we examine each main
evolutionary process in turn: Section \ref{tidalevolution} discusses
tidal evolution timescales and critical semi-major axes pertinent for
satellite spin synchronization, Section \ref{byorp} summarizes current
theories on BYORP evolution and their applicability, Section
\ref{encounters} evaluates if planetary encounters can explain the
eccentric and wide orbits of asynchronous satellites, and Section
\ref{kozai} discusses Kozai resonance and its timescales. We find that
tidal evolution can explain the observed spin-orbital characteristics
of nearly all NEA systems, and we discuss possible evolutionary pathways between
observed spin-orbital states in Section \ref{discussion}. We summarize
this study and its implications in Section \ref{conclusion}.

\subsection{Sample of Well-Characterized Binaries and Triples} \label{sample}

We compile a sample of well-characterized NEA systems (Table
\ref{nea}) that consists of 7 binaries and 2 triples. We will refer to
this sample throughout this paper as we study the spin-orbital origin
of these systems.  This sample includes all NEA systems with known
estimates of system mass, semi-major axis, eccentricity, and component
sizes.  In practice, only systems observed with radar fall in this
class.

Primary and secondary sizes of these
NEAs are obtained from radar shape modeling (when available), or from
range extents estimated from radar images. System masses are derived
from orbital solutions, which are computed based on measurements of
range and Doppler separations.  For triple systems, individual mass
estimates are obtained through dynamical 3-body orbital fits.  For
some binaries, the masses of the primary and of the secondary have
been directly estimated from the observed reflex motion. For all
others, the system mass is apportioned to the individual components by
using size ratios and a common density assumption.

\begin{figure}[htb]
	\centering
	\includegraphics[scale=0.35]{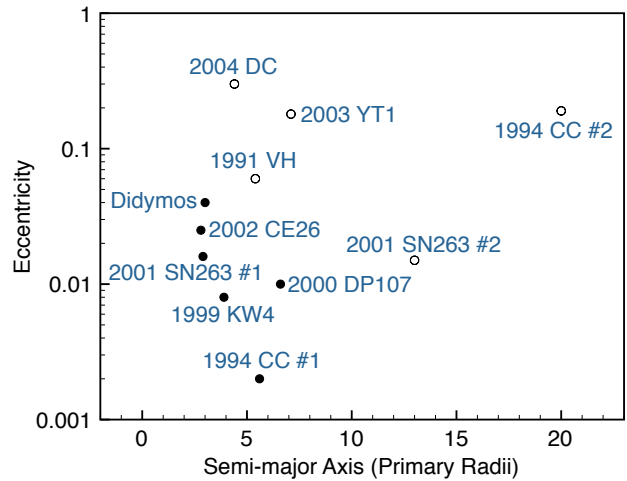}
	\caption{This plot shows the well-characterized NEA systems in
	eccentricity and semi-major axis (in primary
	radius) space, whose orbital parameters are taken from Table
	\ref{nea}. Asynchronous satellites are represented by 
	unfilled circles and synchronous satellites are marked with filled circles. 
	\label{neasample} }
\end{figure}

The semi-major axes and eccentricities of these well-characterized
NEAs are obtained through orbital fits to radar astrometry, and we
list them in Table \ref{nea} and plot them in Figure \ref{neasample}. 
\citet{broz11} have reported a list of asynchronous satellites that
are rapidly rotating, which includes the following satellites in our
sample of NEA systems: 2003~YT1, 1991~VH, 2004~DC, and the outer 
satellites of 2001~SN263 and 1994~CC. These asynchronous satellites
are marked with asterisks in the first column of Table \ref{nea}
and represented by unfilled circles in Figure \ref{neasample}. In this sample,
asynchronous rotators include all satellites with adopted
eccentricities greater than 0.05 and all satellites with semi-major
axes greater than 7 primary radii. These correlated spin-orbital
properties are likely due to the decreasing effects of tidal
dissipation (which can synchronize the satellite's spin to its
orbital period and circularize orbits) at larger semi-major axes.
The exploration of evolutionary processes that can explain these
spin-orbital characteristics will be discussed in the bulk of this
paper.

Plausible 1$-\sigma$ uncertainties for semi-major axes and
eccentricities are compiled from published values (when available) and
listed in Table \ref{nea} along with the adopted values. When
published uncertainties are not available, we obtain uncertainties by
determining their 1$-\sigma$ confidence levels
\citep[e.g.][]{cash76,pres92}.  For each considered parameter, we hold
it fixed at a range of values while simultaneously fitting for all
other parameters.  Since only one parameter is held fixed at a time,
a 1$-\sigma$ confidence region is prescribed by the range of solutions
which exhibit chi-square values within 1.0 of the best-fit chi-square.

All of these well-characterized NEA systems are observed by radar and
have varying degrees of observational coverage and quality.  We have
high confidence in the datasets for binaries 1999~KW4, 2000~DP107, and
1991~VH as well as triples 2001~SN263 and 1994~CC. These systems have
extensive, high signal-to-noise measurements on $\sim$10 epochs over
$\sim$2 weeks, or have been observed on 2 separate apparitions. We
have moderate confidence for 2002~CE26, Didymos, 2004 DC, and 2003
YT1, which all have $\sim$50$-$150 measurements over
4$-$14 days.  These high and moderate confidence datasets comprise the
well-characterized sample of NEA binaries and triples listed in Table
\ref{nea}.  We also mention another radar-observed binary designated
1998~ST27 with an asynchronously rotating satellite \citep{benn03} 
due to its uniquely large separation.  1998~ST27 has a low confidence
dataset with fewer than 40 measurements and inconsistencies in orbit
solutions, and thus we can only determine a lower bound on its
semi-major axis of $\gtrsim$12 primary radii or $\gtrsim$5 km. Its actual
semi-major axis may be much higher. Its large separation makes 1998~ST27
the widest NEA binary discovered so far.

Since all of these NEA systems in Table \ref{nea} are characterized by
radar data, this sample is biased towards binaries and triples
discovered through radar techniques, which typically require close
approaches with Earth of $\lesssim$ 0.1 AU. Close approach data,
including the most recent approach near the time of radar observations
and the current minimum orbital intersection distance (MOID) with
Earth, are listed in Table \ref{nea}. The MOID describes the minimum
distance between two elliptical orbits and disregards the positions of
the bodies in their orbits \citep{sita68}, and is valid as long as the
osculating orbital elements approximate the actual orbits. These
orbital elements certainly change over long periods of time and
therefore close approaches to Earth less than the current MOID could
have occurred in the past.

\subsection{Dynamical and Collisional Lifetimes} \label{lifetimes}

NEAs are short-lived and have dynamical lifetimes on the order of a
few million years \citep{bott02}. These average lifetimes represent how long
they can survive in near-Earth space before plunging into the Sun,
getting ejected from the Solar System, or colliding into a planet.
Due to these short NEA lifetimes, the near-Earth population is continually
replenished by small ($\lesssim$10 km) MBAs. These small MBAs migrate
into near-Earth space through unstable main belt regions that are
permeated by strong resonances with Jupiter and Saturn. Resonance
capture is enabled by radiation forces and collisions, which bring MBAs
into these unstable main belt regions. While in the main belt, these
asteroids have collisional lifetimes that are dependent on their sizes.
For example, an asteroid 500 meters in radius will have a collisional lifetime of 
$\sim$3.8 $\times$ 10$^8$ years \citep{bott05}.

Clearly, collisional lifetimes (while in the main belt) are much greater 
than dynamical lifetimes (while in near-Earth space). It is 
quite plausible that many observed NEA systems had their satellites
formed while still in the main belt, and so their total lifetimes as a
binary or triple will be dominated by their prior collisional
lifetime. An important implication of binary/triple formation in the
main belt is that some evolutionary processes have a longer period of
time within which they may occur. The examination of these
evolutionary processes constitute the bulk of this paper.

\subsection{Binary and Triple Formation} \label{formation}

Binary and triple asteroids form through the generally accepted
rotational fission model \citep{marg02,prav06,wals08}.  The likely
spin-up mechanism is the thermal YORP effect \citep{rubi00}, which
causes mass shedding from the primary.
In this subsection, we introduce the post-fission
dynamics model of \citet{jaco11b}, and in particular we evaluate its
relevance in explaining any of the observed spin-orbital
characteristics of NEA binaries and triples.

\citet{jaco11b} describe the immediate ($\lesssim$1000 years) dynamics
following rotational fission, which include
YORP (to spin up the initial body, and this is the only time a
non-gravitational force is incorporated in their model), secondary
fission (a satellite is rotationally accelerated and then fissions to
create another satellite), tri-axial gravitational potentials, tides,
and solar gravitational perturbations.  The immediate result after
initial fissioning of 
the primary is a chaotic binary, and subsequent evolutionary processes
are mainly determined by mass ratio (mass of secondary/mass of
primary).

\begin{figure}[htb]
	\centering
	\includegraphics[scale=0.35]{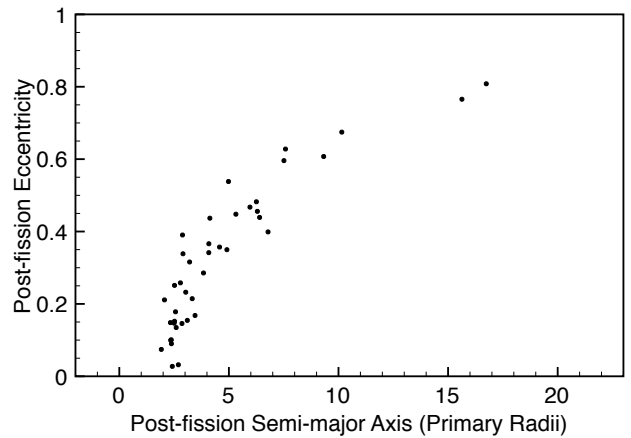}
	\caption{Results from post-fission simulations by \citet{jaco11b} are
	shown after 1000 years of evolution (Seth Jacobson, personal communication,
	2011).
	\label{postfissionae} }
\end{figure}

Binaries with high mass ratios ($>$ 0.2) do not undergo secondary
fission and instead experience tidal dissipation to become doubly
synchronous. This may lead to a contact binary state if BYORP
contracts the orbit. Low mass ratio binaries will mostly disrupt
unless they are allowed to experience secondary fission, which creates
an initially chaotic triple system.  The chaotic triple can stabilize
and become a binary
by ejection or collision of a satellite, which can lower the system's
angular momentum and energy. Resulting binaries that are stable after
1000 years of evolution are shown in Figure \ref{postfissionae}
\citep{jaco11b}.

In Figure \ref{postfissionae}, we point out two scenarios of interest.
First, early tidal evolution can lead to a commonly observed type of
binary: a synchronous satellite with a separation of $\sim$4 primary
radii and an eccentricity of $\sim$0.3$-$0.4 that will continue to
damp by tides. Second, some satellites have large separations (up to
$\sim$10$-$16 primary radii) from their primary with correspondingly
large eccentricities (up to $\sim$0.6$-$0.8). These wide
primary-satellite pairs may explain observed satellites located at
large semi-major axes, although the eccentricities predicted by the
post-fission dynamics model are larger than any observed value.

Triple formation can occur through secondary fission or another round
of YORP-induced primary fission. \citet{jaco11b} include secondary
fission in this model and their simulations produce no stable triple
systems after 1000 years of evolution. They do not model additional
rounds of YORP-induced primary fission, which remains a plausible
explanation for the formation of triple systems. We add the hypothesis
that triples form by first creating a wide binary (such as those seen
in post-fission simulations) through primary fission, and then a
closer-in satellite is formed during a subsequent round of primary
fission. This possibility is also supported by observations of wide
binary 1998~ST27 that has a separation ($\gtrsim$12 primary radii)
consistent with the outer satellite's separation ($\sim$13 primary radii) in
triple 2001~SN263.

Other studies discussing spin-up fissioning either do not mention or
only provide scant information about the resulting spin-orbital
parameters of a newly-formed satellite
\citep[e.g.][]{sche07,hols10,wals08}. The fissioning model of
\citet{wals08} does not attempt to simulate post-fission dynamics
and so does not include tidal interactions. In their simulations,
when satellites grow to 0.3 primary radii, their
separations are 2$-$4.5 primary radii and their eccentricities are
$<$0.15. This range of eccentricities is lower than that found by
\citet{jaco11b} and shown in Figure \ref{postfissionae}.

In essence, the immediate ($\lesssim$1000 years) dynamics ensuing from
this formation scenario provide a pathway for the creation of wide
binaries such as 1998~ST27 and potentially the outer satellites in
triple systems. However, the eccentricities from post-fission dynamics
are too high compared to observed eccentricities, and some 
of the simulated binaries have small semi-major axes of $\sim$2
primary radii that are smaller than any observed separation. Moreover,
the spin states of just-formed satellites will be
asynchronous. 
Clearly, if multiple systems are formed by rotational fission and
follow the post-fission dynamics model of \cite{jaco11b}, there will
be additional processes that evolve the systems, and the exploration
of these processes constitute the bulk and remainder of this paper.

\section{Tidal Evolution} \label{tidalevolution}

In this section, we investigate evolutionary processes due to tidal
effects, and whether tides can explain the observed spin-orbital
characteristics of satellites in NEA systems. The fastest tidal
evolution process (in the absence of other spin-modifying forces) is
the synchronization of a satellite's spin to its orbital period due to
the tides raised on the secondary by the primary
\citep{gold63,gold09}. After synchronization, tidal evolution
continues by modifying the semi-major axis and eccentricity of the
mutual orbit. It is a competing process between the opposing effects
of tides raised on a primary (which increase both the semi-major axis
and eccentricity) and those raised on the secondary (which cause
negligible changes in the semi-major axis, but decrease
eccentricity). Tidal evolution ends after the orbit has circularized
and the primary's spin has synchronized to the mutual orbit period,
resulting in a doubly synchronous system.

Models of tidal evolution for asteroids are governed by two dimensionless
parameters: the effective rigidity $\tilde{\mu}$ and the tidal
dissipation factor $Q$. The non-dimensional effective rigidity
$\tilde{\mu}$ is a function of the body's internal properties such as
density $\rho$ and radius $R$, and is defined as follows for a
monolith: $\tilde{\mu} = (19 \mu) / (2 g \rho R)$ where $g$ represents
self-gravity and $\mu$ is the body's rigidity or shear modulus with
units of pressure \citep{murr99}. The effective rigidity is related to
the tidal Love number $k$, where $k = 1.5 / (1 + \tilde{\mu}) $ for a
homogeneous solid body.  The tidal dissipation factor $Q$ is a quality
factor defined as $Q = (2 \pi E_0) / (\Delta E)$, which describes the
body's effectiveness at dissipating energy \citep{murr99}. If we
consider the body's response to tidal oscillations as a harmonic
oscillator, $E_0$ represents the peak energy stored during a cycle and
$\Delta E$ is the energy dissipated over a cycle. Unfortunately, for
small asteroids there are substantial uncertainties in crucial
parameters $\tilde{\mu}$ (or $k$) and $Q$, in addition to other poorly
known effects, such as the frequency dependence of these two
quantities, their applicability to porous bodies, or even our ability
to capture tidal processes with two idealized numbers.

\subsection{Eccentricity Evolution}

Post-fission eccentricities (Section \ref{formation}) are significantly higher
than observed eccentricities for some NEA systems, and so here we
explore the effects of tides in modifying eccentricity. Several models
have been considered to facilitate calculations of tidal evolution in
asteroids.

In the {\em monolith} model, asteroids are idealized as uniform bodies
with no voids.  In this idealization, the effective rigidity
$\tilde{\mu}_{\rm mono}$ is inversely proportional to the square of
the asteroid's radius $R$: $\tilde{\mu}_{\rm mono}$ $\propto$ $R^{-2}$
or $k_{\rm mono} \propto R^2$.  To arrive at numerical values,
\citet{gold09} used the Moon's radius and Love number $k$ of
$\sim$0.025 \citep{will08} and obtained $\tilde{\mu}_{\rm mono}$
$\sim$ $2 \times 10^8$ (1~km/$R$)$^2$.  The monolith tidal model has
been used to estimate the relative strengths of the components in the
Kalliope-Linus binary system. Since its mutual orbit is found to be
near-circular, comparison of the relative rates of eccentricity
excitation and damping constrains the relative values of
$\tilde{\mu}Q$ for the primary and secondary: \citet{marg03} found
that $\tilde{\mu}Q$ for the secondary is smaller than that of the
primary.

In the {\em rubble pile} model, asteroids are idealized as
gravitational aggregates, i.e.\ composed of smaller elements held
together by gravity only. This assumption is motivated by the low
observed densities of NEA systems and the observed ``spin
barrier''~\citep{prav02}. We will consider two separate rubble pile
models.

\citet{gold09} propose that the relationship between a rubble pile's
effective rigidity $\tilde{\mu}_{\rm rubble}$ and a monolith's
effective rigidity $\tilde{\mu}_{\rm mono}$ of comparable composition
and size is simply $\tilde{\mu}_{\rm rubble}$ $\sim$ $10
\sqrt{\tilde{\mu}_{\rm mono}}$. Thus, this model \citep{gold09}
defines a rubble pile's effective rigidity as $\tilde{\mu}_{\rm
  rubble}$ $\sim$ $10^5 \sqrt{2}$ (1~km/$R$) or Love number $k_{\rm
  rubble} \sim 10^{-5}$ ($R$/1~km). If we assume a common density and
tidal quality $Q$ factor between the primary and secondary, the rubble
pile model adopted by \citet{gold09} gives the ratio of the rates of
eccentricity excitation to damping as 19/28 (irrespective of component
sizes) for a system with a synchronized secondary.  Therefore, in this
model the eccentricity will likely damp for such systems. In
cases where there are density or $Q$ differences between the
components, it is possible the eccentricity can grow.

\citet{jaco11} find that the \citet{gold09} model agrees reasonably
well for data of binaries with primary radii of $\sim$2 km, but not
for systems with very small primary radii. This discrepancy can be
resolved using a different Love number dependence on size, and
\citet{jaco11} describe a power law fit to the logarithmic data of
known synchronous binaries compiled by \citet{prav06}. They find that
$\tilde{\mu}_{\rm rubble} \sim 6 \times 10^4$ ($R$/1~km) or $k_{\rm
  rubble} \sim 2.5 \times 10^{-5}$ (1~km/$R$). Thus, if we apply this
model for binary components of common density and tidal quality factor
$Q$, we find that the ratio of the rates of eccentricity excitation to
damping is ($19{R_s}^2)/(28{R_p}^2$), where $R_p$ is the radius of the
primary and $R_s$ is the radius of the secondary. Thus, eccentricity
will also likely damp in this model.

For all models, the eccentricity evolution 
for a synchronous satellite's orbit \citep{gold63,gold09} is
\begin{align} \label{tides}
	\dfrac{de}{dt} = \dfrac{57}{8}\dfrac{k_p}{Q_p}\dfrac{M_s}{M_p}\left(\dfrac{R_p}{a}\right)^5ne
	- \dfrac{21}{2}\dfrac{k_s}{Q_s}\dfrac{M_p}{M_s}\left(\dfrac{R_s}{a}\right)^5ne
\end{align}
where there are two competing terms corresponding to eccentricity
excitation and damping, respectively. This equation is a function of
tidal Love number $k$, tidal quality factor $Q$, mass $M$, radius $R$,
semi-major axis $a$, mean motion $n$, and eccentricity $e$. The subscripts $p$ and $s$
represent the primary and secondary, respectively. 

Using Equation \ref{tides}, we calculate the circularization
timescales $\tau_{\rm damp}$ of all orbits in our NEA sample given
their current orbital and physical parameters (Table \ref{nea}). These
timescales are presented in Table \ref{tides_ecc}. In our
calculations, we use an assumption of $Q$ $\sim$ 10$-$$10^2$, which is
reasonable for small rocky bodies.
We calculate $\tau_{\rm damp}$ according to both rubble pile models:
the $k \propto R$ model of \citet{gold09} and the $k \propto$ $R^{-1}$
model of \citet{jaco11}. 
With these assumptions, we obtain $\tau_{\rm damp}$ by calculating the
1/$e$ damping timescale (here $e$ is Euler's constant) for
eccentricity by numerically integrating Equation~\ref{tides}.

For the cases of 1999~KW4, 2001~SN263~\#1, 1994~CC~\#1, and
1994~CC~\#2, the tidal Love number model of \citet{gold09} gives a
larger $de/dt$ for the excitation term than for the damping term
(which occurs for these cases due to density differences between the
primary and satellite), and so their eccentricities would
theoretically be predicted to increase. However, most of these
satellites are observed to have circular orbits, which indicates that
either the densities are incorrect, there are $Q$ differences between
the primary and satellite, and/or the tidal model is incorrect.  For
these cases, we suspect that the satellite densities are in error.  If
we assume that all components have a density equal to that of the
primary when we compute $de/dt$, then we find that their
eccentricities will damp as predicted by theory.

The eccentricity damping timescales calculated from Equation
\ref{tides} and presented in Table \ref{tides_ecc} show that there are
differences of several orders of magnitude between timescales
calculated from the two different tidal models of \citet{gold09} and
\citet{jaco11}.  As expected, satellites with larger semi-major axes
such as the outer satellites in 2001~SN263 and 1994~CC will have
mutual orbits that take longer to circularize, and closer-in systems
will have shorter damping timescales. The total possible lifetimes of
each system, which includes collisional and dynamical lifetimes, are
also listed in Table \ref{tides_ecc}. For 1994~CC's outer satellite,
its orbit has a damping timescale that is clearly greater than its
total possible lifetime. 
If this satellite formed from a post-fission dynamics
scenario~\citep{jaco11b} with a high post-fission eccentricity
($\sim$0.8), the system cannot evolve to the currently observed
eccentricity of 0.19 by tides within its lifetime; this remains true
even if there are $Q$ differences between the primary and secondary
(i.e.~let $Q_p=100$ and $Q_s=10$) or if the satellite's size is
underestimated \citep[its radius can be as large as 55 meters;][]
{broz11}. If 1994~CC's outer satellite is to be explained by
post-fission dynamics and tides, then either fissioning can deliver a
far-flung satellite with
lower eccentricities or the equations that idealize tidal interactions
are not sufficient.  For 1999~KW4, 2002~CE26, Didymos, 1991~VH, and
2001~SN263~\#1, their ranges of damping timescales are less than their total
possible lifetimes and this suggests that they should be close to
circular, which is corroborated by the fact that their observed $e$
are all less than 0.1. For most NEAs, we cannot draw firm conclusions
since these calculations are very dependent on crucial parameters such
as $k$ and $Q$, which are poorly constrained for small asteroids;
their damping timescales can be either greater or less than their
total lifetimes.

To summarize, in this subsection we have considered the effect of
tides in modifying eccentricity. Current models show that tides will
damp eccentricity, but increasing eccentricity is possible if the
primary and secondary have different density and/or $Q$
values. 
The post-fission dynamics model of \citet{jaco11b} tends to produce
high-$e$ in distant satellites, and we find that for all satellites
except the outer satellite of 1994~CC, it is possible that tides can
damp their post-fission eccentricities to observed eccentricities
within a collisional lifetime.

\def\arraystretch{1.4}
\begin{deluxetable*}{lrrrrr}
\tablecolumns{6}
\tablecaption{Tidal Timescales for Eccentricity Damping \label{tides_ecc}}
\startdata
\hline \hline

\multicolumn{1}{c}{System} &
\multicolumn{1}{c}{$a$/$R_p$} & 
\multicolumn{1}{c}{$e$} &
\multicolumn{1}{c}{$\tau_{\rm lifetime}$ (yr)} &
\multicolumn{2}{c}{$\tau_{\rm damp}$ (yr):} \\
\multicolumn{1}{c}{} &
\multicolumn{1}{c}{} & 
\multicolumn{1}{c}{} &
\multicolumn{1}{c}{} &
\multicolumn{1}{c}{($k \propto R$)\footnotemark[1]} & 
\multicolumn{1}{c}{($k \propto$ $R^{-1}$)\footnotemark[2]} \\

\hline
2000 DP107 		& 6.6 	& 0.01	& 3.36 $\times$ 10$^8$ & $1 \times 10^8 - 1 \times 10^9$ & $4 \times 10^5 - 4 \times 10^6$	\\			
1999 KW4		& 3.9 	& 0.008	& 4.32 $\times$ 10$^8$ & $3 \times 10^6 - 3 \times 10^7\ddag$& $3 \times 10^4 - 3 \times 10^5$	\\	
2002 CE26		& 2.8 	& 0.025	& 7.03 $\times$ 10$^8$ & $1 \times 10^7 - 1 \times 10^8$ & $4 \times 10^4 - 4 \times 10^5$	\\	
*2004 DC  		& 4.4 	& 0.30	& 2.19 $\times$ 10$^8$ & $2 \times 10^8 - 2 \times 10^9$ & $3 \times 10^4 - 3 \times 10^5$	\\	
*2003 YT1  		& 7.1	& 0.18	& 3.94 $\times$ 10$^8$ & $1 \times 10^9 - 1 \times 10^{10}$ & $1 \times 10^6 - 1 \times 10^7$	\\	
Didymos			& 3.0 	& 0.04	& 3.36 $\times$ 10$^8$ & $5 \times 10^6 - 5 \times 10^7$ & $4 \times 10^3 - 4 \times 10^4$	\\	
*1991 VH 		& 5.4 	& 0.06	& 4.12 $\times$ 10$^8$ & $2 \times 10^7 - 2 \times 10^8$ & $2 \times 10^5 - 2 \times 10^6$	\\	
2001 SN263 \#1 	& 2.9 	& 0.016	& 6.06 $\times$ 10$^8$ & $3 \times 10^6 - 3 \times 10^7\ddag$& $4 \times 10^4 - 4 \times 10^5$	\\	
*2001 SN263 \#2	& 13 	& 0.015	& 6.06 $\times$ 10$^8$ & $4 \times 10^8 - 4 \times 10^9$ & $4 \times 10^7 - 4 \times 10^8$	\\			
1994 CC \#1		& 5.6 	& 0.002	& 2.96 $\times$ 10$^8$ & $5 \times 10^8 - 5 \times 10^9\ddag$& $1 \times 10^6 - 1 \times 10^7$	\\		
*1994 CC \#2 	& 20	& 0.19	& 2.96 $\times$ 10$^8$ & $5 \times 10^{12} - 5 \times 10^{13}\ddag$& $2 \times 10^9 - 2 \times 10^{10}$

\enddata
\tablenotetext{}{
For each system's satellite, we list the adopted values for the observed
semi-major axis $a/R_p$ and eccentricity $e$, its total possible lifetime
$\tau_{\rm lifetime}$ (collisional lifetime plus dynamical lifetime) \citep{bott05}, 
and the eccentricity damping timescale $\tau_{\rm damp}$ due to tides for 
synchronous satellites. We include two tidal models with a different size 
dependence of the Love number $k$. For each model, there is a range of 
values due to adopted values for the tidal dissipation factor $Q$ $\sim$ 
10$-$$10^2$.  \\
* Known asynchronous satellites \\
$\ddag$ Systems in which we assumed that all components had density equivalent to that of primary
}  
\footnotetext[1]{\citet{gold09}}
\footnotetext[2]{\citet{jaco11}}
\end{deluxetable*}

\subsection{Satellite's Spin Evolution}

We consider a satellite's spin evolution due to tidal torques (on the
tide-generated bulge as well as on the permanent
deformation) and the radiative YORP effect.  Spin evolution is
important in our investigation on the origin of the observed spin
states of NEA satellites as well as its consequences for spin-orbit
synchronization-dependent processes such as BYORP.

In the absence of other perturbations, spin-orbit synchronization
of the satellite is the fastest tidal process \citep{gold09} according to
\begin{align} \label{spineqn}
	\dfrac{d\omega_s}{dt} = 5 \pi {\rm sgn}(n - \omega_s) \dfrac{k_s}{Q_s} G \rho_p \left( \dfrac{\rho_p}{\rho_s} \right) \left( \dfrac{R_p}{a} \right)^6
\end{align}
where $\omega$ is the spin rate, $n$ is the mean motion, $k$ is the
Love number, $Q$ is the tidal dissipation factor, $G$ is the
gravitational constant, $\rho$ is the density, $R$ is the radius,
and $a$ is the semi-major axis. Subscripts $p$ and $s$ denote
the primary and satellite, respectively. All observed asynchronous
satellites in our sample are fast rotators that spin faster than
their orbital motions and so $d\omega_s/dt$ will be negative. We
calculate the synchronization timescales $\tau_{\rm sync}$ to despin
from a breakup period ($\sim$2.33 hours) to its orbital period
according to Equation \ref{spineqn}, and present them in Table
\ref{tides_spin}.  These despinning timescales are shorter than the
total lifetimes for most satellites and are shorter than the
eccentricity damping timescales for all satellites (Table
\ref{tides_ecc}). These synchronization timescales will likely be affected
by YORP, which may help speed up synchronization or slow it down.

\def\arraystretch{1.4}
\begin{deluxetable*}{lrrrrrrrr}
\tablecolumns{9}
\tablecaption{Tidal Timescales and Distances for Spin Synchronization \label{tides_spin}}
\startdata
\hline \hline
\multicolumn{1}{c}{System} &
\multicolumn{1}{c}{$a$/$R_p$} & 
\multicolumn{1}{c}{$e$} &
\multicolumn{2}{c}{$\tau_{\rm sync}$ (yr):} &
\multicolumn{2}{c}{$a_{\rm c,tide}$/$R_p$:} &
\multicolumn{1}{c}{$R_H$ (km)} &
\multicolumn{1}{c}{$a_{\rm c,perm}$ (km)} \\
\multicolumn{1}{c}{} &
\multicolumn{1}{c}{} &
\multicolumn{1}{c}{} &
\multicolumn{1}{c}{($k \propto R$)\footnotemark[1]} & 
\multicolumn{1}{c}{($k \propto$ $R^{-1}$)\footnotemark[2]} &
\multicolumn{1}{c}{($k \propto R$)\footnotemark[1]} & 
\multicolumn{1}{c}{($k \propto$ $R^{-1}$)\footnotemark[2]} & 
\multicolumn{1}{c}{} &
\multicolumn{1}{c}{} \\

\hline
2000 DP107 			& 6.55	& 0.01	& 7 $\times$ 10$^6$ $-$ 7 $\times$ 10$^7$ & 6 $\times$ 10$^4$ $-$ 6 $\times$ 10$^5$ & 2.88 $-$ 4.23  & 6.33 $-$ 9.28 & 86 & 1091	 \\
1999 KW4			& 3.86	& 0.008	& 2 $\times$ 10$^5$ $-$ 2 $\times$ 10$^6$ & 4 $\times$ 10$^3$ $-$ 4 $\times$ 10$^4$ & 2.80 $-$ 4.12  & 5.36 $-$ 7.86 & 70 & 1703	 \\
2002 CE26			& 2.78	& 0.025	& 6 $\times$ 10$^4$ $-$ 6 $\times$ 10$^5$ & 6 $\times$ 10$^2$ $-$ 6 $\times$ 10$^3$ & 2.80 $-$ 4.11  & 6.14 $-$ 9.01 & 514& 4584 \\
*2004 DC  			& 4.41	& 0.30	& 3 $\times$ 10$^6$ $-$ 3 $\times$ 10$^7$ & 1 $\times$ 10$^3$ $-$ 1 $\times$ 10$^4$ & 1.41 $-$ 2.06  & 5.24 $-$ 7.65 & 44 & 189 	  \\
*2003 YT1  			& 7.15 & 0.18	& 1 $\times$ 10$^7$ $-$ 1 $\times$ 10$^8$ & 4 $\times$ 10$^4$ $-$ 4 $\times$ 10$^5$ & 2.67 $-$ 3.93  & 6.62 $-$ 9.71 &113 & 1488 \\
Didymos				& 2.95	& 0.04	& 8 $\times$ 10$^4$ $-$ 8 $\times$ 10$^5$ & 2 $\times$ 10$^2$ $-$ 2 $\times$ 10$^3$ & 2.30 $-$ 3.38  & 6.35 $-$ 9.33 & 109& 891 	 \\
*1991 VH 			& 5.43	& 0.06	& 1 $\times$ 10$^6$ $-$ 1 $\times$ 10$^7$ & 3 $\times$ 10$^4$ $-$ 3 $\times$ 10$^5$ & 3.40 $-$ 5.00  & 6.38 $-$ 9.37 & 105& 1987	 \\
2001 SN263 \#1 		& 2.92	& 0.016	& 1 $\times$ 10$^5$ $-$ 1 $\times$ 10$^6$ & 2 $\times$ 10$^3$ $-$ 2 $\times$ 10$^4$ & 3.40 $-$ 4.99  & 6.46 $-$ 9.48 & 343& 5415  \\
*2001 SN263 \#2		& 12.8 & 0.015	& 7 $\times$ 10$^7$ $-$ 7 $\times$ 10$^8$ & 8 $\times$ 10$^6$ $-$ 8 $\times$ 10$^7$ & 5.16 $-$ 7.58  & 7.43 $-$ 10.9 & 343& 5532  \\
1994 CC \#1			& 5.58	& 0.002	& 2 $\times$ 10$^7$ $-$ 2 $\times$ 10$^8$ & 3 $\times$ 10$^4$ $-$ 3 $\times$ 10$^5$ & 2.03 $-$ 3.00  & 6.16 $-$ 9.06 & 86 & 911 	 \\
*1994 CC \#2 		& 19.8 & 0.19 &3 $\times$ 10$^{10}$ $-$ 3 $\times$ 10$^{11}$&2 $\times$ 10$^7$ $-$ 2  $\times$ 10$^8$& 1.71 $-$ 2.52  & 5.81 $-$ 8.55  & 86 & 553 	
\enddata
\tablenotetext{}{
For each system's satellite, we list the adopted values for the observed semi-major
axis $a$/$R_p$ and eccentricity $e$, the tidal spin synchronization timescale $\tau_{\rm sync}$
starting from the breakup rate
in the absence of other effects such as YORP, the critical semi-major axis
$a_{\rm c,tide}$/$R_p$ within which an initially asynchronous satellite can achieve
synchronization against the effects of YORP torques that compete with tidal torques,
Hill radii $R_H$, and an upper limit to the critical semi-major axis
$a_{\rm c,perm}$ at which an initially synchronous satellite would break spin lock 
due to the effects of YORP. The critical semi-major axes are calculated using
f$_Y$ $\sim 5 \times 10^{-4}$. When applicable, we include 2 tidal models with a
different size dependence of the Love number $k$. For each model, there is a range 
of values due to adopted values for the tidal dissipation factor $Q$ $\sim$ 10$-$$10^2$. \\
* Known asynchronous satellites}  
\footnotetext[1]{\citet{gold09}}
\footnotetext[2]{\citet{jaco11}}
\end{deluxetable*}

Accordingly, we investigate the effect of YORP during a satellite's spin
evolution. For both cases of an initially asynchronous or
synchronous satellite, we ask, how does a satellite's rotation evolve
when the semi-major axis changes?
We do not assume that the orbit migration is dominated by any 
particular mechanism. Perturbations such as tides or BYORP can increase 
a satellite's semi-major axis, and BYORP can also decrease the semi-major
axis.

\subsubsection{Initially Asynchronous Satellite}

First, we examine an initially asynchronous satellite, perhaps newly
formed. All satellites will go through this stage.  Its spin evolution
will be affected by YORP as well as the torque due to the satellite's
tidal bulge. The restoring torque due to the satellite's permanent
deformation is not applicable in this case because this torque
averages out unless the satellite is in a spin-orbit resonance
\citep{murr99}.

We apply the torque equations due to tides raised on the satellite 
by the primary and YORP using the formalism given in \citet{murr99} and
\citet{stei11} for the magnitudes of the torques $N$:
\begin{equation}
	N_{\rm tide} = \frac{3}{2}\frac{k_s}{Q_s}\dfrac{G M_p^2 R_s^5}{a^6} \label{tidetorque}
\end{equation}
\begin{equation}
	N_{\rm YORP} = \dfrac{R_s^3 |f_Y| L_{\odot}}{6 c d_{\odot}^2 \sqrt{1 - e_{\odot}^2}} \label{yorptorque}
\end{equation}
where $G$, $c$, and $L_{\odot}$ are the gravitational constant, the
speed of light, and the solar luminosity, respectively.  The
heliocentric parameters include semi-major axis $d_{\odot}$ and
eccentricity $e_{\odot}$.  Tidal parameters include the Love number
$k_s$ and dissipation factor $Q_s$, where the subscript $s$ denotes
that these quantities are for the satellite. $M_p$ is the primary's
mass, $a$ is the semi-major axis, and $R_s$ is the satellite's radius.
Following \citet{gold09}, we include a YORP reduction factor $f_Y$
that can be positive or negative, and accounts for a reduction from
its maximum possible value. This factor is necessary since the
incoming radiation is not completely absorbed and reemitted
tangentially along the satellite's equator.

The tidal torque will try to establish synchronization between the
secondary's spin and its orbital period, and the YORP torque will
perturb the secondary's spin in either direction. Their relative
contributions are dependent on the semi-major axis. 
If the torques act in the same direction, the satellite's spin will
synchronize with its orbital period. If the torques are competing,
spin synchronization is not guaranteed.  The critical semi-major axis
$a_{\rm c,tide}$ at which the magnitudes of these torques are equal is
\begin{equation}
	a_{\rm c,tide} = \left(3 R_s M_p d_{\odot}\right)^{1/3} \left(\dfrac{k_s G c \sqrt{1 - e_{\odot}^2}}{Q_s |f_Y| L_{\odot}}\right)^{1/6} \label{crita}
\end{equation}
where the constants and variables are the same as defined for
Equations \ref{tidetorque} and \ref{yorptorque}.

Satellites with orbital distances less than this critical semi-major
axis $a_{\rm c,tide}$ will become synchronous, whether or not YORP and
tides act in the same direction.  For orbital distances greater than
the critical semi-major axis, satellites will likely remain
asynchronous if tides and YORP act in opposite directions.  We
calculate $a_{\rm c,tide}$ for all systems in our sample (Table
\ref{nea}), using $Q_s = 10-100$ and $f_Y \sim 5 \times 10^{-4}$ as
suggested by observations of (54509) YORP (formerly 2000~PH5)
\citep{lowr07,tayl07}. 
A list of $a_{\rm c,tide}$ is given in Table \ref{tides_spin}.
Uncertainties in the tidal Love number $k_s$ and dissipation value
$Q_s$ produce a range of critical semi-major axes per NEA binary in
our sample, and different values for $f_Y$ would result in an even
broader range.

We now discuss the agreement between our calculations of $a_{\rm
  c,tide}$ and the observed spin states (given in
Table~\ref{tides_spin}) of NEA satellites in our sample.  Two
satellites, 2002 CE26's secondary and 2001 SN263's inner satellite,
have current semi-major axes less than the range of possible $a_{\rm
  c,tide}$ values calculated in this table, and they are both observed
to be synchronous. Two satellites, the outer satellites of 2001 SN263
and 1994 CC, have current semi-major axes larger than their range of
possible $a_{\rm c,tide}$ values, and they are both observed to be
asynchronous; this suggests that tides and YORP act in opposing
directions for these two satellites. All other satellites whose
current semi-major axes may or may not be larger than their $a_{\rm
  c,tide}$ include synchronous and asynchronous rotators. 
The dominance of YORP for satellites such as 2001~SN263~\#2 explains
how it could have a tidally-circularized orbit yet be asynchronous.

Further observations of asynchronous satellites can help constrain
which tidal model \citep{gold09,jaco11} best captures the behavior of
small bodies, assuming that tides and YORP are the dominant processes
affecting satellite spin states.  For this test we focus on
asynchronous satellites because in that case the balance of tidal and
YORP torques requires their observed semi-major axes to be larger than
their computed $a_{\rm c,tide}$ values.  Comparison of the observed
$a$ to the predicted $a_{\rm c,tide}$ for a number of asynchronous
satellites of different sizes may reveal which tidal model is more
appropriate.  We encourage observations of asynchronous satellites to
enable this test.  Unfortunately the same test cannot be applied to
the more numerous synchronous satellites because their spin can be
explained by tides and YORP torques acting in the same direction with
the observed $a$ $<$ $a_{\rm c,tide}$ or $a$ $>$ $a_{\rm c,tide}$, or
by tides and YORP acting in opposite directions with $a$ $<$ $a_{\rm
  c,tide}$.

\subsubsection{Initially Synchronous Satellite}

We now consider the case of an initially synchronous satellite.
Its spin evolution will be affected by YORP, the satellite's tidal
bulge, and additionally, a restoring torque due to the satellite's
permanent deformation. The permanent quadrupole moment of the
satellite plays a role for spin states in spin-orbit resonance and is
thus applicable when considering the spin evolution of a synchronous
satellite.  The expression for the magnitude of the torque due to
permanent deformation is given by \citet{murr99} as
\begin{equation} \label{perm}
N_{\rm perm} = \dfrac{3}{2} (B-A)\dfrac{GM_p}{a^3} \sin(2\psi)
\end{equation}
where $B$ and $A$ are the satellite's equatorial principal moments of
inertia, $G$ is the gravitational constant, $M_p$ is the mass of the
primary, $a$ is the semi-major axis, and $\psi$ is the amplitude of
the libration.

We compare the tidal torque $N_{\rm tide}$ to the permanent
deformation torque $N_{\rm perm}$ for the case of 1999 KW4, which is
currently the only NEA binary with published shape information of the
secondary. Using 1999 KW4's $(B-A)$ value for its secondary, $\psi
\sim \pi/4$ (maximum amplitude), values for $Q$ from 10 to 100, and
both tidal models \citep{gold09,jaco11}, we find that $N_{\rm perm}$
dominates over $N_{\rm tide}$ by at least six orders of
magnitude. Therefore, for the case of an initially synchronous satellite
that we examine here, we will consider only the permanent deformation
torque (and not the torque on the tidal bulge) as well as the YORP
torque.  The permanent deformation torque will seek to maintain
spin-orbit synchronization, and the YORP torque will attempt to spin
the satellite in either direction away from synchronization.  Balance
of the permanent deformation and YORP torques yields a critical
semi-major axis $a_{\rm c,perm}$ given as
\begin{equation} \label{crita2}
a_{\rm c,perm} = \dfrac{1}{R_s} \left(\dfrac{9 (B-A) G M_p \sin(2\psi) c d_{\odot}^2 \sqrt{1 - e_{\odot}^2}}{|f_Y| L_{\odot}} \right)^{1/3}
\end{equation}
where the constants and variables are the same as defined for 
Equations \ref{yorptorque} and \ref{perm}.

In the absence of significant eccentricities, this critical semi-major
axis $a_{\rm c,perm}$ separates regions where synchronization can be
maintained and where synchronization can be broken.  We seek an upper
limit for $a_{\rm c,perm}$ to determine if synchronization can be
maintained to distances as large as the Hill radius, which is a
requirement for synchronization-dependent processes such as BYORP to
create asteroid pairs. In Table \ref{tides_spin}, we present
calculations of an upper limit $a_{\rm c,perm}$ for all satellites in
our sample. We adopt a maximum libration amplitude $\psi \sim \pi/4$
and calculate $(B-A)$ for each satellite by scaling from the $(B-A)$
of one of the most elongated NEAs known, Geographos \citep{ostr95}:
\begin{equation}
	(B-A)_{\rm x} = \dfrac{(B-A)_{\rm Geo}}{M_{\rm Geo} R_{\rm Geo}^2} M_{\rm x} R_{\rm x}^2
\end{equation}
where subscripts $x$ and Geo represent the considered satellite
and Geographos, respectively. The mass is $M$ and the
equivalent radius (if it were spherical) is $R$.
The equivalent radius of Geographos is $\sim$1.28 km
\citep{huds99}, and if we assume a typical rubble pile
density of 2 g cm$^{-3}$, its mass is $\sim$1.8 $\times$ 10$^{13}$ kg.
Using a triaxial ellipsoid assumption to calculate
its moments of inertia, we find $(B-A)_{\rm Geo}/(M_{\rm Geo}R_{\rm Geo}^2)$
to be $\sim$0.56. Similarly as we did earlier for calculations of $a_{\rm c,tide}$,
we adopt $f_Y \sim 5 \times 10^{-4}$. These values are used in
the calculation of an upper limit $a_{\rm c,perm}$ shown
in Table \ref{tides_spin}.

Table \ref{tides_spin} shows that for all NEA systems, $a_{\rm c,perm}$
is much larger than the Hill radius. This suggests that once a satellite
has synchronized its spin to its orbital period, it can remain so
out to far distances even in the presence of YORP. Therefore, all
observed asynchronous satellites (2004~DC, 2003~YT1, 1991~VH, and the
outer satellites of triples 2001~SN263 and 1994~CC) have probably 
never been synchronous unless they experienced a planetary encounter that
disrupted their synchronous spins. 

An alternative proposal was recently put forth by \citet{jaco11c}, who
suggest that longitude librations of the satellite will occur about a
direction that is not aligned with the line connecting the central
body and the satellite.  They hypothesize that this angular offset becomes
increasingly significant as the orbit expands (due to tides and
BYORP), and eventually results in breaking a satellite's spin lock.
If this model is correct, then synchronization cannot be maintained to
far distances.  However this model seems to produce substantial
angular offsets only for satellites with very small moment of inertia
differences ($(B-A)/C$), and therefore does not appear to be effective
for the overwhelming majority of satellites.

To summarize, we find that a newly post-fissioned, asynchronous
satellite can become synchronous within its lifetime and this can
explain all observed synchronous and circular binaries and inner
satellites of triples. Asynchronous outer satellites in both triples
are unable to synchronize by tidal torques because of the larger
influence of YORP, and this may also be the explanation for
asynchronous binaries 2004~DC, 2003~YT1, and 1991~VH.  The dominance
of YORP over tides at large distances may explain why the outer
satellite in 2001~SN263 remains asynchronous despite having an orbit
that appears to have tidally circularized.

\subsection{Semi-major Axis Evolution}

The post-fission dynamics model of \citet{jaco11b} shows how
satellites can be deposited at large separations up to $\sim$16 primary radii
(Section \ref{formation}) and can explain the outer satellite in
2001~SN263 (at $\sim$13 primary radii) and perhaps also the outer
satellite in 1994~CC (at $\sim$20 primary radii). Here we investigate
if another mechanism, namely tidal evolution, can bring satellites
from closer-in to wide orbits.

To explore this possibility, we study three cases of wide orbits:
2001~SN263~\#2, 1994~CC~\#2, and we also consider a hypothetical wide
binary modeled after 1998~ST27, which has a primary radius of 0.42 km
and a lower limit on its observed semi-major axis of 5 km.  For these
three systems, we calculate the tidal timescales for the semi-major
axis to increase from one primary radius to its observed value. For
both tidal models \citep{gold09,jaco11} and using $Q =$ 10$-$100, we
find that the tidal timescales for such increases in semi-major axis
are $\sim$10$^{9}$$-$10$^{10}$ years for 2001~SN263~\#2,
$\sim$10$^{10}$$-$10$^{13}$ years for 1994~CC~\#2, and
$\sim$10$^{9}$$-$10$^{11}$ for the hypothetical wide binary modeled
after 1998~ST27. The total possible lifetimes of these satellites
(see Table \ref{tides_ecc}; for the hypothetical binary, its 
lifetime is $\sim$3.4 $\times$ 10$^8$ years) are
shorter than their tidal expansion timescales. This suggests that
tides cannot account for the wide orbits of these satellites, if the
tidal model is correct. Other influencing factors such as ``tidal
saltation,'' the idea of mass lofting from the primary and briefly
entering into orbit to transfer orbital angular momentum to the
satellite before falling back down to the primary's surface, may speed
up tidal expansion \citep{harr09}.

Next, we investigate if tides can explain the wide orbits by using
different assumptions of interior properties. Assuming tidal
dissipation $Q$ values of 10$-$100, we calculate their material
properties in order for their semi-major axes to increase from one
primary radius to the currently observed semi-major axis within their
NEA-specific lifetimes. We find that in order for tides to be responsible for the increase in
semi-major axis within the considered system's lifetime, we will need to invoke
the following values for $k_p$: 
$\sim$0.00003$-$0.00030 for 2001~SN263~\#2, $\sim$0.0054$-$0.0545 for
1994~CC~\#2, and $\sim$0.00028$-$0.00285 for the hypothetical
binary modeled after 1998~ST27. The required $k$ value for 1994~CC~\#2
is prohibitive and is almost as large as or larger than the Moon's $k$
of 0.025 \citep{will08}.

In summary, tidal evolution cannot explain the semi-major axes of
widely-separated systems without invoking unusual material
properties,
lower $Q$ values than assumed, a different tidal model, or a
combination of these factors.
Our analysis here supports the idea that post-fission dynamics 
may be largely responsible for some of the wide orbits observed 
in NEA systems.

\section{BYORP Evolution} \label{byorp}

The BYORP effect occurs for a synchronously rotating satellite with an
asymmetrical shape. A synchronous satellite has permanent leading and
trailing hemispheres, and an asymmetrical shape will result in
re-radiation of absorbed sunlight that is uneven between the two
hemispheres.  This disparity results in a net acceleration (or
deceleration) of the satellite's orbit and can therefore cause orbital
evolution.  This effect has not been observationally verified, and in
this section we evaluate BYORP's relevance in explaining the observed
spin-orbital characteristics of NEA systems by introducing current
theoretical models predicting BYORP's effects and timescales.

For all observed asynchronous satellites, we rule out BYORP as a major
player in their recent evolution because this effect depends on
spin-orbit synchronization; without synchronous spin, the effect
cancels out. This is applicable to nearly half of the satellites
in our sample that are asynchronous (Table \ref{nea}): 2004~DC,
2003~YT1, 1991~VH, 2001~SN263~\#2, and 1994~CC~\#2.
We now examine scenarios in which BYORP may have had an important role
in the past evolutionary histories of these systems, assuming their
satellites used to be synchronous. 

First, we consider the case where a
synchronous satellite evolved via BYORP and broke spin-lock through a
planetary encounter. Such a scenario would imply that its past
BYORP-affected orbital properties would be erased in part by the
scattering encounter; therefore, any observed properties cannot be
wholly attributed to BYORP, but would also be attributed to the flyby.

Second, we consider the case where BYORP expands a synchronous
satellite's orbit and increases the eccentricity enough to break
spin-lock and result in chaotic asynchronous rotation \citep{cuk10}.
This could perhaps explain the observed eccentric and asynchronous
satellites, but we find this unlikely for two reasons. First,
synchronous re-capture is thought to occur rapidly \citep[][on the
  order of $\sim$10$^3$ years]{cuk10} 
and would prevent a substantial population of asynchronous binaries from forming.  
Second, this process would not explain how all observed
asynchronous satellites acquired spin periods much less than their
respective orbital periods. 

In essence, although we cannot rule out that BYORP played a role in
the evolution of asynchronous satellites, we find that tides (for spin
synchronization) or planetary encounters are required to explain the
data, whereas BYORP is not.  Therefore, we find that BYORP alone
cannot explain the properties of asynchronous satellites, which
includes all satellites with large semi-major axes.  We apply the same
reasoning to widely-separated systems such as 1998~ST27 and the outer
satellites of the triples.  Because neither tides alone nor BYORP
alone can readily explain the origin of large semi-major axes, we
conclude that these properties may be primarily a result of
post-fission dynamics rather than evolutionary processes.

For all remaining synchronous satellites, 
BYORP may be responsible for their observed spin-orbital characteristics.
However, BYORP is neither necessary nor sufficient to explain these
properties: another evolutionary process (tides) is required to first
synchronize the satellite's spin before BYORP can operate.  BYORP's
relevance for NEA systems is also complicated by the short BYORP
timescales. 
An important unresolved issue with our understanding of BYORP has to
do with conflicting models on the sign of $de/dt$ with respect to
$da/dt$.  These issues are discussed in the following paragraphs.

Current theories predict different behaviors of semi-major axis and
eccentricity evolution due to BYORP
\citep{gold09,cuk10,mcma10b,mcma10a,stei11}.  \citet{mcma10b} describe
how the overall orbital evolution over long timescales due to BYORP
causes the semi-major axis and eccentricity to evolve in {\em opposite
  directions}, even with the inclusion of libration effects.
\citet{cuk10} also include the effects of a secondary's librations due
to its elongated shape and predict that librations dominate over
direct perturbations by BYORP.  They describe how the semi-major axis
and eccentricity evolve in the {\em same direction}. They suggest that
the most likely outcome of an initially expanding orbit (with
$\dot{e}>0$ in their model) is chaotic rotation of the secondary
followed by synchronous spin re-establishment and inward migration,
which would prevent evolution to large semi-major axes.
\citet{stei11} also find that the preferred end state for BYORP is
shrinkage of the semi-major axis. On the other hand, \citet{mcma10c}
suggest that librational motion for binaries like 1999~KW4 
\citep[$\sim$5$^{\circ}$ amplitude librations;][]{ostr06}
will remain small for expanding orbits (with $\dot{e}<0$ in their model), 
preventing chaotic rotation of the secondary, and leading to large 
semi-major axes. Future measurements of spin states of satellites can
help constrain which model best captures the correct behavior.

The timescales for BYORP evolution are thought to be fast.
\citet{cuk05} find that for most asteroid shapes, the BYORP torque is
significant and can modify the satellite's orbital semi-major axis and
eccentricity on timescales of $\sim$10$^5$ years. Similar expansion
timescales have been found by \citet{mcma10b}, who assert that orbits
can typically expand to their Hill sphere due to BYORP on the order of
10$^4$ $-$ 10$^6$ years. Currently, 1999~KW4 is the only
well-characterized NEA binary with published shapes of both primary
and secondary, and \citet{mcma10b} predict an orbit expansion rate of
7 cm yr$^{-1}$ (a prediction corroborated by \citet{stei11}) and an
orbit-doubling time of $\sim$22,000 years. Under the influence of
BYORP, primary oblateness, and solar perturbations, they predict that
1999~KW4 will reach its Hill radius in $\sim$54,000
years. Incorporation of librational effects into the model by
\citet{mcma10b} causes a longer BYORP evolution, but the mutual orbit
will still expand to its Hill radius for small librational motion 
to at least 12$^{\circ}$ \citep{mcma10c}.

There are potential puzzles with the hypothesized short BYORP
timescales and assertions that BYORP-induced orbit expansion can reach
the Hill radius.  First, binary formation would need to be rapid
enough to produce the observed fraction of NEAs with satellites.
Since NEAs have dynamical lifetimes on the order of a few million
years \citep{bott02}, much larger than the BYORP disruption timescale
of $\sim$0.1 million years, the binary production rate would need to
match the rate of binaries disrupted by BYORP for a steady-state
binary NEA population. Part of this issue (at least for nominally half
of all systems in which BYORP causes inward migration) may be
mitigated if a stable equilibrium exists between BYORP and tides
\citep[e.g][]{jaco11}.  If BYORP acts to contract the orbit and tides
cause expansion of the orbit, there is a critical semi-major axis at
which their effects may balance and result in a stable equilibrium.
Second, according to \citet{mcma10b}, BYORP can expand the orbits
(nominally half) of satellites to their Hill radii. This means we
should observe some systems that are at significant fractions of their
Hill radii; for instance, the Hill radii of these systems are
typically a few hundred primary radii. However, we do not observe any
systems wider than 20 primary radii. Moreover, such BYORP-induced
expansion to the Hill sphere will cause the expanding binary to become
more susceptible to planetary encounters by shortening the encounter
timescale and strengthening the perturbative effects of
encounters. Using equations and timescales in \citet{fang11i}, we
calculate that if 1999~KW4 reaches a separation halfway to its Hill
radius, Earth encounters at a typical encounter velocity of 12 km
s$^{-1}$ are expected to disrupt the binary at encounter distances of
$\sim$28 Earth radii, which can occur every $\sim$260,000 years. With
a separation of 90\% of its Hill radius, disruptive encounters can
occur every $\sim$100,000 years. As a result, we suggest that
planetary encounters may also create asteroid pairs in the near-Earth
population.

We also mention that the predictions for 1999~KW4's expansion to its
Hill radius given by \citet{mcma10b} are not ruled out by our
calculations of upper limit critical semi-major axes $a_{\rm c,perm}$
at which a synchronous satellite will have its spin lock broken.  We
calculated these critical semi-major axes (Table \ref{tides_spin})
using Equation \ref{crita2}, which compares the torque due to a
satellite's permanent deformation (which maintains synchronization in
the absence of significant eccentricities)
and the torque due to YORP (which can spin an asteroid away from
synchronization). Table \ref{tides_spin} shows that for all NEA
satellites, the critical semi-major axis $a_{\rm c,perm}$ is much
larger than the Hill radius, meaning that synchronization can be
maintained in the absence of other perturbations.  Thus, if the
predictions by \citet{mcma10b} are correct (i.e.\ eccentricities remain
small when the orbit expands), the formation of asteroid pairs by 
BYORP-induced orbit expansion (and possibly planetary encounters) is possible. 

To summarize, we find that BYORP is not currently acting for nearly
half of the satellites in our sample of NEA binaries and triples
because they are asynchronously rotating.  We also find difficulties
associated with the hypothesis that BYORP operated in the past
evolutionary histories of these currently asynchronous satellites,
although we cannot rule out that their orbital properties may have
been shaped in part by BYORP.  For the remaining synchronous
satellites, BYORP may be acting but is not required
to explain observed orbital properties.

\section{Evolution by Close Planetary Encounters} \label{encounters}

In this section, we examine if close planetary encounters can explain
the observed spin-orbital properties of well-characterized NEA
systems.

The majority of satellites in our sample of NEA systems are
synchronously rotating and on near-circular orbits.  This includes
2000~DP107, 1999~KW4, 2002~CE26, Didymos, and the inner satellites of
2001~SN263 and 1994~CC.  We find that close planetary encounters are
unlikely to explain these properties.  We expect that planetary encounters
are strong enough to cause changes to an asteroid's spin state 
\citep[e.g.][]{sche00,sche04} and orbital eccentricity
\citep[e.g.][]{fang11i}, such that we would not expect a majority of
synchronous and circular binaries if planetary flybys were a dominant
process.
Similarly, planetary flybys are also not likely to explain the
spin-orbital state of 2001~SN263's outer satellite, which although
asynchronous, has a circular orbit.

For asynchronous binaries 2003~YT1, 2004~DC, and 1991~VH, their
eccentricities are 0.18, 0.3, and 0.06, respectively. The outer
satellite of 1994~CC is also asynchronous with an orbital eccentricity
of 0.19. 2003~YT1, 2004~DC, and 1991~VH have semi-major axes that
indicate their expected eccentricities following post-fission dynamics
were higher or comparable to their current eccentricities (Figure
\ref{postfissionae}).  
If they formed at their current semi-major axes, either tides did not
have sufficient time to damp the eccentricities to circular orbits
(Table \ref{tides_ecc}), or a planetary encounter may have erased some
of the tidal damping effects by increasing the eccentricities.
If they formed closer-in (with correspondingly lower eccentricities
following post-fission dynamics), a planetary encounter may have
increased both their semi-major axis and tidally-damped eccentricity
to observed values.
1994~CC's outer satellite has tidal damping timescales much longer than its
total possible lifetime, and so its current eccentricity is too low to
be explained by post-fission dynamics and tides. In this case, it
is possible that planetary encounters may have lowered its
eccentricity from a high predicted value following post-fission
dynamics to its observed value.

To investigate these scenarios, we employ results from a companion
paper on the effect of planetary encounters with binary asteroids
\citep{fang11i}.  We find that for a planetary flyby to increase each
of the asynchronous binaries' eccentricities from a tidally-damped
value of 0 to its observed value
would take $\sim$4.94 Myr for 2003~YT1, $\sim$6.54 Myr for 2004~DC,
and $\sim$0.56 Myr for 1991~VH. These encounter timescales represent
the average time for an eccentricity increase when the binary is in
near-Earth space, assuming the NEA follows a trajectory from main belt
source regions
to its current orbit in near-Earth space. These timescales 
can occur within the near-Earth dynamical lifetime on the order of a
few million years. Inclusion of additional planets and repeat passes
will make it more likely that planetary encounters can affect orbital
properties (see \citet{fang11i} for details).  For the case of 1994~CC
with a starting eccentricity of 0.8, the largest decreases in
eccentricity happen for shorter encounter distances. However, at
shorter encounter distances, unstable encounters with collisions and
ejections dominate. For instance, at an encounter distance of 2 Earth
radii, the average post-encounter eccentricity is $\sim$0.58 (not low
enough to match the observed eccentricity of 0.19) and the percentage
of stable encounters is less than 1\%.  Smaller encounter distances
are even more problematic. Scenarios with repeat passes, where the
eccentricity has a net decrease to its observed value, are also very
unlikely. Therefore, our results suggest that planetary encounters
cannot decrease the orbital eccentricity of 1994~CC's outer satellite
from high values to moderate values.

Next, we investigate if the widest orbits observed in NEA systems
originate from post-fission dynamics (Section \ref{formation} and
Figure \ref{postfissionae}) or possibly another mechanism, namely
planetary encounters. We examine a hypothetical binary modeled after
1998~ST27 with a large separation of $\sim$16 primary radii and an
eccentricity of 0.3. Our simulations used encounter distances of
2$-$12 $R_{\Earth}$ and encounter velocities of 8$-$24 km s$^{-1}$ for
an Earth-mass perturber. We find that encounters cannot create wide
systems, except very rarely.  Here are a few illustrative cases we
examine:

1. We perform simulations starting with a circular binary with a
separation of 4 primary radii, which is typical for NEA binaries. For
a typical encounter velocity of 12 km s$^{-1}$, encounters at a
distance of 8 $R_{\Earth}$ showed that $\sim$7\% of stable encounters
at least doubled in semi-major axis and none of the stable binaries
quadrupled in semi-major axis. Encounters resulting in stable binaries
that at least doubled in semi-major axis had post-flyby eccentricities
of $\gtrsim$ $0.45$, which is larger than any observed eccentricity
but may damp within a dynamical lifetime.

2. We perform a ``repeat encounter'' scenario starting with an
eccentric binary ($e =$ 0.5, 0.7) with a separation of 8 primary
radii. These initial conditions assume that an a prior encounter already
doubled the semi-major axis from 4 to 8 primary radii and increased
the eccentricity to high values. We find that it is possible to both
decrease the eccentricity and further increase the semi-major axis to
observed values. However, this only occurred in $\sim$5\% of stable
encounters with a fine-tuned initial semi-major axis and eccentricity
in order to match the observed values.

3. We investigate if this hypothetical wide binary used to be a triple
system by performing simulations of a primary asteroid and 2
satellites initially located at 4 and 16 primary radii with circular
orbits. The chance of ejecting the inner satellite with the outer
satellite intact is possible, but only occurs in $\lesssim$3\% of all
unstable encounters with encounter distances less than
$\sim$8$R_{\Earth}$.  When it does happen, the final eccentricity of
the outer satellite is typically higher than any observed
eccentricity.

Our work shows that planetary encounters are not necessary to explain
the observed spin-orbital properties of binary NEAs, but in reality,
close planetary encounters will occur and they will change the orbital
characteristics of NEA multiples \citep{fang11i}. To reconcile these
two results, there are a few possibilities: (a) the observed
synchronous and circular systems are more recent migrants to
near-Earth space that have not yet undergone deep planetary encounters, 
and oppositely, the asynchronous and eccentric systems have been 
interacting with planets for longer periods of time, 
or (b) the observed systems have encountered
terrestrial planets, but tidal evolution occurs on a faster timescale,
and has managed to synchronize and circularize some
systems. Possibility (b) may or may not be supported by calculations
of tidal damping timescales (Table \ref{tides_ecc}) depending on the
tidal Love number model and NEA-specific encounter timescales.

In summary, we find that planetary encounters cannot explain observed
synchronous satellites nor nearly circular orbits. For asynchronous
satellites with non-circular orbits (2004~DC, 2003~YT1, and 1991~VH),
we find that planetary encounters are a plausible explanation that can
reproduce their observed eccentricities, although not necessary to
explain the data (observed properties can be explained by tides alone).
However, planetary encounters cannot decrease 1994~CC's eccentricity
from a high post-fission dynamics value to the observed value. We also
find that the observed wide orbits were unlikely to be reached through
planetary encounters.

\section{Evolution by Kozai Resonance} \label{kozai}

Now we examine orbital evolution by the Kozai resonance \citep{koza62}
by reviewing its effects and timescales as well as its applicability
to observed NEA systems.

Kozai resonance \citep{koza62} is a secular effect causing angular
momentum exchange between an inner body's orbital eccentricity and
relative inclination with a massive, outer perturber. In this
resonance, eccentricity $e$ and inclination $I$ are coupled, and the
quantity $\sqrt{1 - e^2}\cos I$ is conserved in the idealistic case of
an outer perturber with a circular orbit. For an initially circular
binary, large Kozai oscillations can occur if the relative
inclination between the inner and
outer orbits is at least $\sim$39.2 degrees, and this critical
inclination depends on the ratio of their semi-major axes. Since this
mechanism can result in high eccentricities, we consider Kozai
resonance between an NEA binary and the Sun as the outer perturber
\citep{pere09}. Triple systems are not considered here, as
\citet{fang11} have shown that the presence of a second satellite 
damps any Kozai oscillations in the system and thus will not be a
relevant mechanism in modifying the satellites' eccentricities.

For binaries under the influence of the Kozai mechanism, we can
estimate the Kozai period $P_{\rm Kozai}$ or typical oscillation 
timescale between limiting values of $e$ and $I$ as \citep{kise98}
\begin{equation} \label{kozaip}
	P_{\rm Kozai} = \dfrac{2 P_{\odot}^2}{3 \pi P_{\rm binary}} (1 - e_{\odot}^2)^{3/2} \dfrac{M_p + M_s + M_{\odot}}{M_{\odot}}
\end{equation}
where $P_{\odot}$ is the heliocentric orbital period, $P_{\rm binary}$
is the binary's mutual orbital period, and $e_{\odot}$ is the
heliocentric eccentricity. The binary's primary mass is $M_p$, the
secondary mass is $M_s$, and the outer perturber's mass is the Sun's
mass ($M_{\odot}$) in the situation we examine here. In the absence of
other effects, we calculate the Kozai periods for all
well-characterized NEA binaries and present them in Table
\ref{kozaitimes}. 
Binaries with small heliocentric distances and/or large heliocentric
eccentricities (such as 1999~KW4) have short oscillation periods.
The very short timescales in Table \ref{kozaitimes} indicate that
evolution by the Kozai mechanism is faster than any other
evolutionary process examined in this study.

Determination of whether binaries are in the Kozai regime or not can
be assessed for at least 2 near-circular binaries in our sample, 2000~DP107 and
1999~KW4, for which we have detailed information about the orientation
of the mutual orbital plane. For these binaries, we calculate the
current inclination between the binary's mutual orbit and the Sun's
apparent orbit. We find that neither 1999~KW4 nor 2000~DP107 have
inclinations that meet the critical Kozai angle. Other binaries
in our sample do not currently have reliable orbital orientations.
The difficulty of measuring precise orbital plane orientations without
sufficient radar coverage makes the assessment of Kozai influences
difficult to verify at the moment, and additional observations of NEA
binaries are encouraged.

Fulfillment of the required Kozai inclination will be affected by
processes that can change the relative inclination between the
binary's mutual orbit and its heliocentric orbit. For a satellite in
an equatorial orbit with a semi-major axis of several primary radii,
\citet{cuk05} describe a possible equilibrium state where YORP and
BYORP torques balance.  They describe how this scenario can occur when
the inclination between the primary's equatorial plane and the
heliocentric orbit is $\sim$50$-$60 degrees. If the binary's mutual
orbit is in the same plane as the primary's equator (as would be
expected from the generally accepted rotational fission formation
model), then the stable inclination of $\sim$50$-$60 degrees would be
the angle between the binary orbit and heliocentric orbit.
\citet{stei11} describe a different stable inclination that will be
reached under the effects of BYORP, which will tend to orient a
binary's orbit into an inclination of either 0 or 90 degrees relative
to its heliocentric orbit.  However, we note that in the absence of
Kozai-damping processes, a ``stable'' inclination of $\sim$50$-$60 or
90 degrees would cause a binary to undergo Kozai cycles. Although the
value of the stable inclination is unclear due to different
predictions by \citet{cuk05} and \citet{stei11}, we suggest that
radiative torques may cause a binary that is initially not affected by
Kozai effects to enter the Kozai regime when its relative inclination
with the Sun's apparent orbit is sufficiently high.

Next, we consider the effect of primary oblateness (described by a
$J_2$ term) in modulating the Kozai effect for NEA binaries. The
critical semi-major axis $a_{\rm c, J_2}$ for the transition between
the influence regions of primary oblateness and solar dynamics such as
the Kozai resonance is \citep{nich08}
\begin{align} \label{nichoeqn}
	a_{\rm c, J_2} = \left( 2 J_2 \dfrac{M_p}{M_{\odot}} R_p^2 a_{\odot}^3 \right)^{1/5}
\end{align}
where $J_2$ approximates the non-spherical shape of the primary by its
level of oblateness, $M_p$ is the mass of the primary, $M_{\odot}$ is
the mass of the Sun, $R_p$ is the primary's radius, and $a_{\odot}$ is
the heliocentric semi-major axis. We calculate $a_{\rm c, J_2}$ for
all NEA binaries in our sample for $J_2$ values ranging from 0.001 to
0.1, and we find that the range of possible $a_{\rm c, J_2}$ values is
larger than the observed semi-major axis for all binaries. This implies
that these binaries are dominated by primary oblateness, which will
cause orbital precession that can completely suppress Kozai cycles.

Lastly, we note that this mechanism can alter an NEA binary's spin
state and eccentricity, but cannot change the semi-major axis of the
binary's mutual orbit in the averaged problem. Therefore, it cannot
explain the observed wide orbits, which again suggests that these
large separations are inherited from post-fission dynamics.

To summarize, Kozai resonance is not applicable for systems where its
effects may be damped, and this includes all satellites in triple
systems and binaries where primary oblateness may dominate. This
includes all systems in our sample of well-characterized binaries and
triples. The averaged Kozai effect also cannot modify the semi-major
axis.

\def\arraystretch{1.4}
\begin{deluxetable}{lrrr}
\tablecolumns{4}
\tablecaption{Kozai Oscillation Periods \label{kozaitimes}}
\startdata
\hline \hline

\multicolumn{1}{c}{Binary} &
\multicolumn{1}{c}{$a$ (km)} & 
\multicolumn{1}{c}{$e$} &
\multicolumn{1}{c}{$P_{\rm Kozai}$ (yr)} \\

\hline
2000~DP107	& 2.62 & 0.01 & 89.11 \\
1999~KW4	& 2.55 & 0.008 & 10.78 \\
2002~CE26	& 4.87 & 0.025 & 756.60 \\
*2004~DC 	& 0.75 & 0.30 & 269.78 \\
*2003~YT1 	& 3.93 & 0.18 & 58.40 \\
Didymos	    & 1.18 & 0.04 & 545.99 \\
*1991~VH	& 3.26 & 0.06 & 81.34
\enddata
\tablenotetext{}{
For each binary, adopted values for the observed semi-major axis $a$
and eccentricity $e$ as well as the Kozai oscillation period 
$P_{\rm Kozai}$ are listed. \\
* Known asynchronous satellites}
\end{deluxetable}

\section{Combined Origin and Evolution of Spin-Orbital Properties} \label{discussion}

Here we summarize the relevance of each evolutionary process examined in
this study (tides, BYORP, planetary encounters, and Kozai effects)
in the context of explaining the origin of observed spin-orbital properties
of NEA systems in our sample (Table \ref{nea}). Then we discuss
evolutionary pathways between observed spin-orbital states and
disrupted binaries (asteroid pairs and contact binaries).

Table \ref{origin} shows how each considered evolutionary process
matches up against each other in explaining the spin-orbital origin of
binaries and triples in our sample. Tidal evolution is a dominant
process that can explain the satellite's spin state and orbital
eccentricity for nearly all systems assuming formation by rotational
fission followed by post-fission dynamics.  All other examined
processes either depend on tidal evolution and/or are not required to
explain the observed systems.  For instance, BYORP requires a
tidally-synchronized system to operate.
Although BYORP is not required to explain the data, it may be acting
in synchronous systems or may be pivotal in a stable equilibrium state
where the effects of tides and BYORP cancel \citep[i.e.][]{jaco11}.
Flybys, or planetary encounters, can potentially explain several
asynchronous and non-circular systems, but not the majority of
circular and synchronous systems.  The Kozai effect is not applicable
for primary oblateness-dominated NEA systems in our sample.  We also
find that observed wide orbits, such as those of binary 1998~ST27 and
the outer satellites of 2001~SN263 and 1994~CC, are most likely a
direct byproduct of post-fission dynamics~\citep{jaco11b} because none
of the four evolutionary processes as currently modeled seem capable
of delivering satellites to such large separations.

\def\arraystretch{1.4}
\begin{deluxetable*}{lrrllll}
\tablecolumns{7}
\tablecaption{Origin of Spin-Orbital Properties \label{origin}}
\startdata
\hline \hline

\multicolumn{1}{c}{System} &
\multicolumn{1}{c}{$a$/$R_p$} & 
\multicolumn{1}{c}{$e$} &
\multicolumn{1}{c}{Tides} &
\multicolumn{1}{c}{BYORP} &
\multicolumn{1}{c}{Flyby} &
\multicolumn{1}{c}{Kozai} \\

\hline
2000 DP107 		& 6.6 	& 0.01	& \textcolor{ForestGreen}{Yes} & \textcolor{ForestGreen}{Yes} & \textcolor{Red}{No}  & \textcolor{Red}{No} \\
1999 KW4		& 3.9 	& 0.008	& \textcolor{ForestGreen}{Yes} & \textcolor{ForestGreen}{Yes} & \textcolor{Red}{No} & \textcolor{Red}{No} \\
2002 CE26		& 2.8 	& 0.025	& \textcolor{ForestGreen}{Yes} & \textcolor{ForestGreen}{Yes} & \textcolor{Red}{No} & \textcolor{Red}{No} \\
*2004 DC  		& 4.4 	& 0.30	& \textcolor{ForestGreen}{Yes} & \textcolor{Red}{No} & \textcolor{ForestGreen}{Yes} & \textcolor{Red}{No} \\
*2003 YT1  		& 7.1	& 0.18	& \textcolor{ForestGreen}{Yes} & \textcolor{Red}{No} & \textcolor{ForestGreen}{Yes} & \textcolor{Red}{No} \\
Didymos			& 3.0 	& 0.04	& \textcolor{ForestGreen}{Yes} & \textcolor{ForestGreen}{Yes} & \textcolor{Red}{No} & \textcolor{Red}{No} \\
*1991 VH 		& 5.4 	& 0.06	& \textcolor{ForestGreen}{Yes} & \textcolor{Red}{No} & \textcolor{ForestGreen}{Yes} & \textcolor{Red}{No} \\
2001 SN263 \#1 	& 2.9 	& 0.016	& \textcolor{ForestGreen}{Yes} & \textcolor{ForestGreen}{Yes} & \textcolor{Red}{No} & \textcolor{Red}{No} \\
*2001 SN263 \#2	& 13 	& 0.015	& \textcolor{ForestGreen}{Yes} & \textcolor{Red}{No} & \textcolor{Red}{No} & \textcolor{Red}{No} \\
1994 CC \#1		& 5.6 	& 0.002	& \textcolor{ForestGreen}{Yes} & \textcolor{ForestGreen}{Yes} & \textcolor{Red}{No} & \textcolor{Red}{No} \\
*1994 CC \#2 	& 20	& 0.19	& \textcolor{Red}{No} & \textcolor{Red}{No} & \textcolor{Red}{No} & \textcolor{Red}{No} 
\enddata
\tablenotetext{}{
For each system's satellite, we list the adopted values for the observed
semi-major axis $a/R_p$ and eccentricity $e$, as well as which
evolutionary process(es) can explain its observed spin-orbital state.
`Yes' (in green) means that the considered process is plausible and we find no evidence
to rule it out; the opposite response is `No,' (in red) meaning the considered 
process is highly unlikely. \\
* Known asynchronous satellites}  
\end{deluxetable*}

\begin{figure*}[t]
	\centering
	\includegraphics[scale=0.95]{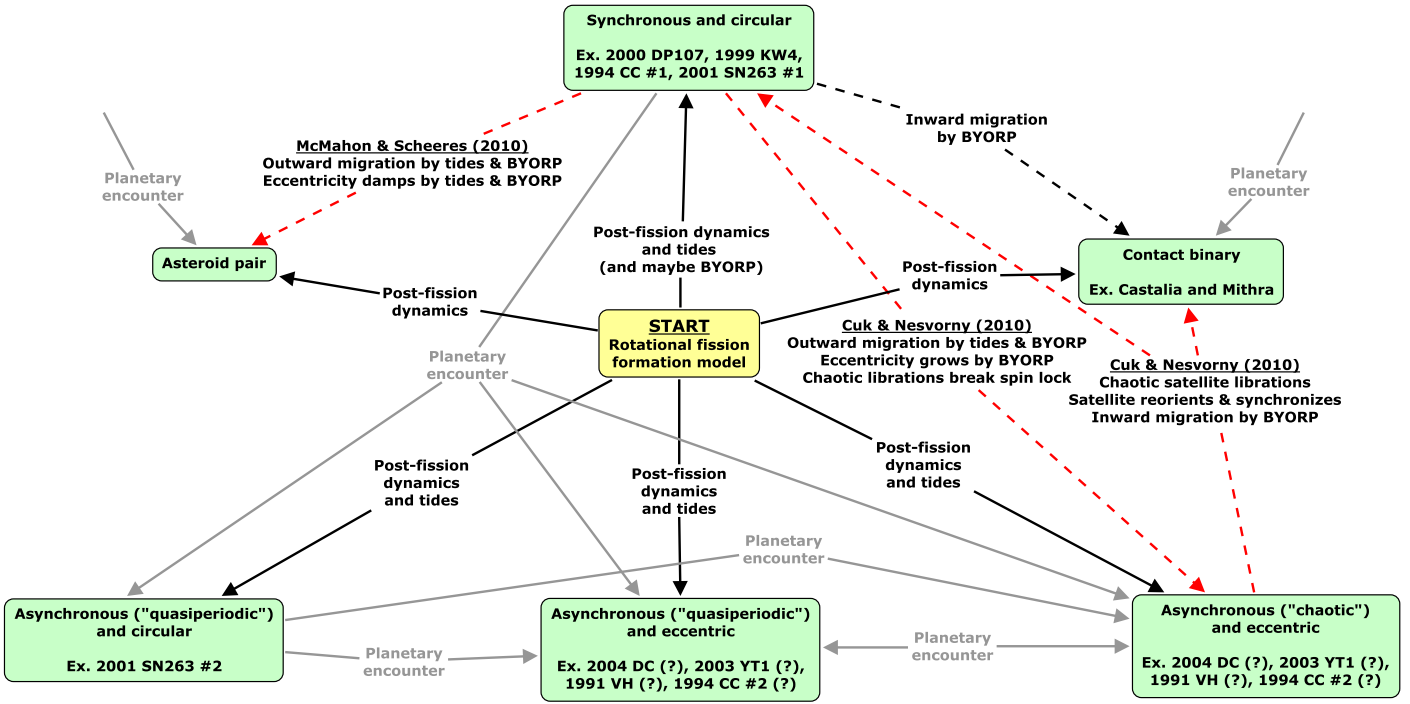}
	\caption{Possible spin-orbital pathways are shown for an
          evolving satellite in a newly post-fissioned NEA system with
          the assumption that Kozai oscillations are not relevant. Pathways
		  that start and end in the same spin-orbital configuration are
		  not shown in this figure, even though they may occur. Two
          different models are given for BYORP and are highlighted in red.
		  \citet{cuk10} predict that BYORP-induced change in eccentricity
		  has the same direction as the change in semi-major axis, and
		  they model how a chaotically asynchronous satellite can reorient
		  and regain synchronization. On the other hand, \citet{mcma10b,mcma10a} 
		  predict that BYORP-induced change in eccentricity has a direction 
		  opposite of the change in semi-major axis. For synchronous and
		  circular systems, it is possible that they are in the stable
		  equilibrium of \citet{jaco11}. See Section \ref{discussion}
          for additional discussion regarding this figure. \label{flow}}
\end{figure*}

Now we discuss these evolutionary processes as an ensemble of pathways
between spin-orbital states and disrupted states such as asteroid pairs and
contact binaries. These pathways are shown in Figure \ref{flow} with
the assumption that these systems are not affected by the Kozai
effect. In the figure, the yellow box is the starting point for a
rotationally fissioning asteroid. Green boxes represent potential
outcomes. The evolutionary pathways are drawn in solid black (for
post-fission dynamics), solid gray (for planetary encounters), and
dotted black (for BYORP), with red lines representing differences
between the two different BYORP models.  \citet{cuk10} predict that
BYORP-induced change in eccentricity has the same direction as the
semi-major axis change (i.e.~when the semi-major axis increases, the
eccentricity increases).  \citet{mcma10b,mcma10a} predict that the
eccentricity moves in the direction opposite to the semi-major axis
change (i.e.~when the semi-major axis increases, the eccentricity
decreases).

As adopted in Figure \ref{flow}, the term {\em asynchronous} can refer
to two types of states where the satellite's spin is non-resonant with
its orbital period: ``quasiperiodic'' rotation, which is stable, and
``chaotic'' rotation, which is unstable.  When these spin state
trajectories are depicted on surface of section plots,
``quasiperiodic'' trajectories are smooth curves and ``chaotic''
trajectories are random points filling in an area on the plot. The
stable or unstable nature of an asynchronous rotation can be readily
discerned on such surface of section plots, as discussed in
\citet{murr99} and shown for Saturn's satellite Hyperion in
\citet{wisd84}. A satellite's principal moments of inertia $(B-A)/C$
determine the strength of spin-orbit resonances.
The higher the eccentricity, the lower the necessary $(B-A)/C$ at
which chaotic rotation can occur. An increase in eccentricity 
(due to perturbations such as BYORP) can lead to overlap between 
spin-orbit resonances. Simultaneous libration in more than
one spin-orbit resonance is not possible, and this leads to chaotic
behavior. A satellite can depart an unstable, ``chaotic'' state by 
reaching a stable ``quasiperiodic'' state and temporarily re-establish
synchronization. In some models, this allows BYORP to regain control 
of the system and evolve it away from the chaotic states, as shown 
in simulations for 1999~KW4-like binaries by \citet{cuk10}.

In Figure \ref{flow}, we discuss spin-orbital pathways that are common to both BYORP
models (lines not shown in red).  Post-fission dynamics 
($<$1000 years) combined with tidal dissipation (and possibly BYORP) 
can produce many observed examples of NEA systems,
including both synchronous and asynchronous rotators, both circular
and eccentric orbits, contact binaries, and asteroid pairs.  Evolution
between spin-orbital outcomes (highlighted in green in the figure) is
possible through close planetary encounters; this can cause both
asynchronous and synchronous rotators in circular orbits to evolve
into asynchronous satellites with eccentric orbits, or any
state into an asteroid pair or contact binary (shown by arrows with 
no starting point in the figure).

We now consider inward evolution
from ``synchronous and circular'' and ``asynchronous
(chaotic) and eccentric'' states. A synchronous satellite will be
affected by BYORP and depending on its shape and rotation, can migrate
inwards and become a contact binary. 
In the model of \citet{cuk10}, a chaotically-spinning (and therefore
asynchronous) satellite in an eccentric orbit will eventually
temporarily regain synchronization with an orientation switched by 180
degrees, in which case BYORP would regain control of the system's
evolution and restore it to a stable state. Inward migration can follow and 
lead to the formation of a contact binary or, after sufficient inward
migration, the effects of BYORP and tides can cancel, leading to a
stable equilibrium \citep[e.g.][]{jaco11} and a tidally-damped
eccentricity. This would result in a synchronous satellite in a circular orbit.

We now consider outward evolution from the
``synchronous and circular'' and
``asynchronous (chaotic) and eccentric'' states.  According to
\citet{cuk10}, outward migration leads to an increase in eccentricity,
so a synchronous and circular state can become chaotically
asynchronous and eccentric. Once chaotically asynchronous and
eccentric, it would remain so unless the satellite can re-orient
itself such that if synchronization is temporarily regained, BYORP can
migrate the satellite inwards.  According to \citet{mcma10b,mcma10a},
outward migration leads to a decrease in eccentricity, which means
that synchronization will be maintained, aided in part by the
restoring torque due to the satellite's permanent deformation. In this
model, outward migration would allow the ``synchronous and
circular'' state to evolve into an asteroid pair.

Examples of NEA satellites in each spin-orbital state are also given in Figure
\ref{flow}.  There are several examples of each
end state, except for asynchronous satellites with circular orbits
and asteroid pairs.
The only known example of an asynchronous satellite with a circular
orbit is the outer satellite in triple 2001~SN263. 
Asynchronous and circular configurations may be rare because
asynchronous satellites are more likely to be well-separated from
their primary, and larger separations are more susceptible to stronger
disruptive planetary encounters with
short encounter timescales \citep{fang11i}.  Strong and frequent
planetary encounters can easily evolve asynchronous and circular
configurations to asynchronous and eccentric states, making it more
rare to observe an asynchronous and circular system.

As for asteroid pairs, none have been definitively reported in the
near-Earth population although they have been observed in the
main belt \citep[i.e.][]{prav10}.
We briefly discuss implications for the formation of NEA asteroid
pairs. Figure \ref{flow} shows how asteroid pairs
can form directly through post-fission dynamics or planetary
encounters. The BYORP model by \citet{mcma10b,mcma10a} provides an
additional route to their creation via BYORP-dominated expansion of
a satellite's orbit and the BYORP model by \citet{cuk10} does not
predict this.
Determination of likely asteroid pair formation
mechanisms can elucidate which BYORP model (\citet{cuk10} or
\citet{mcma10b,mcma10a}) predicts the correct behavior. If
BYORP-induced expansion of a binary's mutual orbit is responsible for
creating NEA asteroid pairs, then we should expect to observe a few
binaries with very large separations unless they are efficiently
disrupted by planetary flybys (see Section \ref{byorp}).
Alternatively, \citet{cuk10} have hypothesized that asteroid pairs are
the result of scattering among satellites in triple systems. Such
scattering can occur if both satellites are or used to be synchronous
and BYORP led them into convergent orbits, which would result in
unstable orbits and scattering. Components would then collide or be
ejected.

\section{Conclusion} \label{conclusion}

Radar observations of well-characterized NEA binaries and triples have
uncovered a diverse set of spin-orbital properties: synchronous and
asynchronously rotating satellites, circular and eccentric orbits, and
a few large separations between the primary and the satellite. The
formation of these satellites by rotational fission followed by a
post-fission dynamics model~\citep{jaco11b} can produce asynchronous
satellites with a variety of primary separations and high orbital
eccentricities.  We investigated how a newly formed system can evolve
to one of the observed systems by evaluating these evolutionary
processes: tides, BYORP, planetary encounters, and Kozai effects.

We found that post-fission dynamics and tides can explain the observed
semi-major axes, eccentricities, and satellite spin states of nearly
all binaries and triples in our sample.  Other evolutionary mechanisms
do not appear to be dominant processes: BYORP is not applicable to
asynchronous systems and is not required to explain the observed data,
even though it may be acting; planetary encounters are likely not
responsible for creating synchronous and circular configurations; and
the Kozai effect will typically be suppressed by the primary's
oblateness.

We also illustrated the evolutionary pathways for satellites in
binaries and triples after they have formed. Evolutionary processes
such as tides, planetary encounters, and BYORP can evolve a system
between synchronous and circular, asynchronous and circular, and
asynchronous and eccentric configurations.

\acknowledgments

We thank Peter Goldreich, Bill Bottke, Matija \'Cuk, and Shantanu Naidu
for useful discussions, Seth Jacobson for providing the results from
his post-fission dynamics simulations, and Jay McMahon for sending us
a copy of his thesis. We are also grateful to the reviewer for helpful
comments. This work was partially supported by NASA Planetary
Astronomy grant NNX09AQ68G.

\bibliographystyle{apj}
\bibliography{binaries}

\end{document}